\documentclass[%
showpacs,
preprintnumbers,
 amsmath,amssymb,
 aps,
prb,
]{revtex4-1}

\usepackage{amsmath}
\usepackage{amssymb}
\usepackage{graphicx}
\usepackage{bm}
\usepackage[pagebackref=false,colorlinks,linkcolor=blue,citecolor=black]{hyperref}
\usepackage[mathlines]{lineno}
\usepackage{siunitx}
\usepackage[
total={6.5in,8.75in}, top=1.2in, left=0.9in, includefoot,
]{geometry}

\newcommand{\x}{exciton}
\newcommand{\xc}{exciton }
\newcommand{\xcs}{excitons }
\newcommand{\FO}{F\"{o}rster }

\begin{document}

\title{Global Equilibrium and Non-Equilibrium Theory of Hopping Exciton Transport in Disordered Semiconductors}

\author{Mehdi Ansari-Rad}
\email{ansari.rad@shahroodut.ac.ir}
\affiliation{Department of Physics, Shahrood University of Technology, Shahrood, Iran}

\author{Stavros Athanasopoulos}
\email{astavros@fis.uc3m.es}
\affiliation{Departamento de F\'isica, Universidad Carlos III de Madrid, Avenida Universidad 30, Legan{\'e}s 28911, Madrid, Spain}

\date{\today}	  

\begin{abstract}
We develop a temperature dependent theory for singlet exciton hopping transport in disordered semiconductors. It draws on the transport level concept within a F\"orster transfer model and bridges the gap in describing the transition from equilibrium to non-equilibrium time dependent spectral diffusion. We test the validity range of the developed model using kinetic Monte Carlo simulations and find agreement over a broad range of temperatures. It reproduces the scaling of the diffusion length and spectral shift with the dimensionless disorder parameter and describes in a unified manner the transition from equilibrium to non-equilibrium transport regime. We find that the diffusion length in the non-equilibrium regime does not scale with the the third power of the F\"orster radius. The developed theory provides a powerful tool for interpreting time-resolved and steady state spectroscopy experiments in a variety of disordered materials, including organic semiconductors and colloidal quantum dots. 
\end{abstract}

\pacs{71.35.Aa, 61.43.Bn, 73.63.Kv, 78.67.Hc, 78.66.Qn, 78.55.Kz}

\maketitle

\section{\label{sec:INTRO}Introduction} 
The phenomenon of exciton diffusion is found to play a role in a remarkably wide range of physical systems including disordered organic semiconductors~\cite{BardeenReview2014,mikhnenko2015exciton}, nanocrystalline quantum dots~\cite{akselrod2014subdiffusive,sensors,ScholesDots,Geissler2016}, semiconducting carbon nanotubes~\cite{Finnie,Kato,Zhao,Hertel} and photosynthetic biological systems~\cite{Sundstrom}. Moreover, there is a growing interest in describing electronic excitation energy transfer because exciton dynamics determines function in many technological applications. For example, in thin film organic solar cells exciton diffusion drives charge separation~\cite{Samuel-review17,Feng18}, in organic light emitting diodes it determines the brightness and color of the device~\cite{ZhaoOLED}, in scintillator detectors it controls the response function and yield~\cite{QDScint}, while in quantum communication systems it facilitates photon antibunching~\cite{HtoonPRL2015}.

In disordered semiconductors that display weak intermolecular interactions, excitations created upon light absorption, carrier recombination or annihilation processes are typically Frenkel excitons that are localized on single chromophore units (molecule, conjugated segment, quantum dot) and have a finite lifetime before relaxation to the ground electronic state occurs by radiative or non-radiative process. In the weak coupling regime, excitons transfer from one unit to the other with a Markovian incoherent hopping process and transport can be described as a simple diffusive motion~\cite{volkhardMay}. However, chromophore units are not equivalent to each other as they can have different on-site excitation energies due to the different local environment, structure or size as well as different excitonic couplings with neighbors. As a consequence, the energy landscape has a distribution that is often approximated by a Gaussian~\cite{kohlerBassler} and the standard deviation of the distribution defines the disorder parameter $\sigma$. Therefore, in the course of time excitations sample the energetic landscape and on average relax to lower energy sites until they 'settle down' to a steady state and equilibrium is achieved. 
However, because excitations have a finite lifetime $\tau$, the relaxation process might be incomplete and consequently the \xc transport out of equilibrium~\cite{athanasopoulos2009PRB}. It should be emphasized that this spectral relaxation process is different from the initial rapid vibronic relaxation~\cite{kerstingVibronicVsRelaxation}. Another consequence of the finite lifetime is that excitations have a limited spatial diffusion range, determined by the diffusion length $L_D$~\cite{JPCCtrap,athanasopoulos2009PRB,Rorich}. Spectroscopic techniques such as time-resolved and time-integrated fluorescence spectroscopy can provide information on spectral diffusion~\cite{Menke,Lin2014,fennel2012forster} and a number of organic systems have been studied over a range of temperatures~\cite{Meskers01,Gaab-Bardeen,Herz04,madigan2006modeling,mikhnenko2008temperature,hoffmann2010spectral}.

A common misconception exists, that in practical device applications at room temperature, equilibrium transport prevails and the description of transport in terms of normal diffusion is sufficient. However, the distinction between equilibrium and non-equilibrium exciton transport is quite a subtle one and the transport regime is not uniquely defined only by temperature.  
Whilst significant progress has been made on understanding temperature dependent spectral relaxation and exciton diffusion, including experimental measurements~\cite{Fayer90,Meskers01,Gaab-Bardeen,Herz04,mikhnenko2008temperature,hoffmann2010spectral,lin2015temperature} and computational models~\cite{Owens89,Fayer90,Fleming92,athanasopoulos2009PRB,athanasopoulos2013Hoffmann,HowCharge,Papadopoulos,JPCCtrap,EvgueniaPRL,JPCLintra,bjorgaard2015simulations,Bardeen2007OrientationalEffects}, currently there is no analytical theory that can describe the transition from equilibrium to non-equilibrium transport. In contrast, for \textit{charges} it has been suggested that the transport problem can be modeled as a multiple-trapping process and it has been shown that a unique level in the energy distribution exists, the transport energy (TE), that plays the same role as the mobility edge in the multiple-trapping mechanism~\cite{grunewald1979hopping,baranovskii1995concept,baranovskii1997applicability}. Note, that in contrast to the long-range nature of the dipole-dipole interaction facilitating singlet exciton transport~\cite{valeur2012molecular}, charge transport in disordered semiconductors occurs via a short-range tunneling mechanism~\cite{baranovskii2014theoretical}.

In this paper we shall develop and test a theory that can treat the dynamics of exciton diffusion at both the equilibrium and non-equilibrium transport regime. In what follows, we develop a formalism based on the TE concept for the calculation of singlet exciton transport parameters, such as relaxation energy and diffusivity, including their temporal dependence. The general formalism is described in section II, while section III includes the main results (parts A, B, D) along with a comparison of the theory to Monte Carlo simulations (part C). Section IV summarizes the work and draws conclusions.

\begin{figure}[!t]
\scalebox{1}{\includegraphics{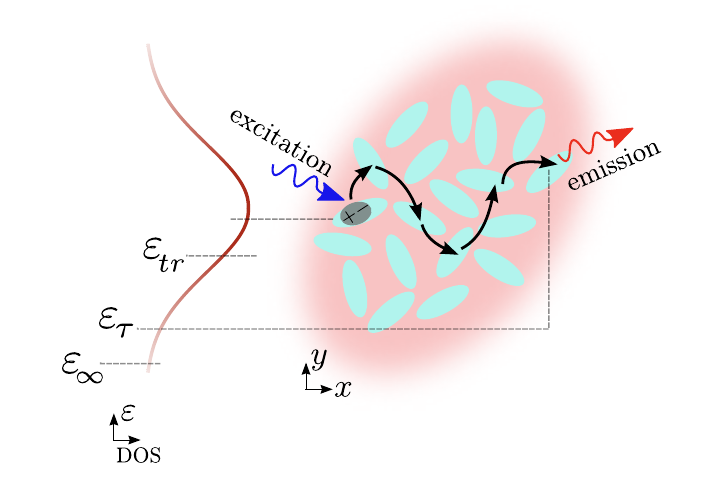}}
\caption{ \label{fig:1}\small \baselineskip=0.59cm
Schematic illustration of interacting units in a disordered semiconductor, resulting to a Gaussian broadened excitonic DOS. Singlet \xc diffusion via F\"{o}rster-type energy transfer process, triggers energy relaxation towards the equilibrium energy $\varepsilon_{\infty}$. Due to the finite lifetime, excitons may decay at a higher energy, $\varepsilon_{\tau}$. $\varepsilon_{tr}$ is the transport energy level. 
}
\end{figure}

\section{\label{sec:TE}Transport Energy level for \FO transfer} 
We consider thermally assisted \FO energy transfer between localized states described by the rate\cite{movaghar1986diffusion,athanasopoulos2009PRB}

\begin{eqnarray}
\nu(\varepsilon_d\rightarrow \varepsilon_a) = \frac{1}{\tau}  S(R)
\exp\left[- \frac{\Delta \varepsilon+\vert \Delta \varepsilon \vert }{2k_{\mathrm{B}}T} \right]
\label{eq:Forster}
\end{eqnarray}
with 
\begin{eqnarray}
S(R) = \left( \frac{R_{\mathrm{F}}}{R} \right)^6
\label{eq:R6}
\end{eqnarray}
where $\tau$ is the intrinsic \xc lifetime, $R_{\mathrm{F}}$ is the \FO radius, determined by the donor-acceptor spectral overlap, and $k_{\mathrm{B}}T$ is the thermal energy. $\Delta \varepsilon = \varepsilon_a - \varepsilon_d$ is the difference between the donor and acceptor energies and $R$ is the corresponding distance. 

\begin{figure}[!t]
\scalebox{1}{\includegraphics{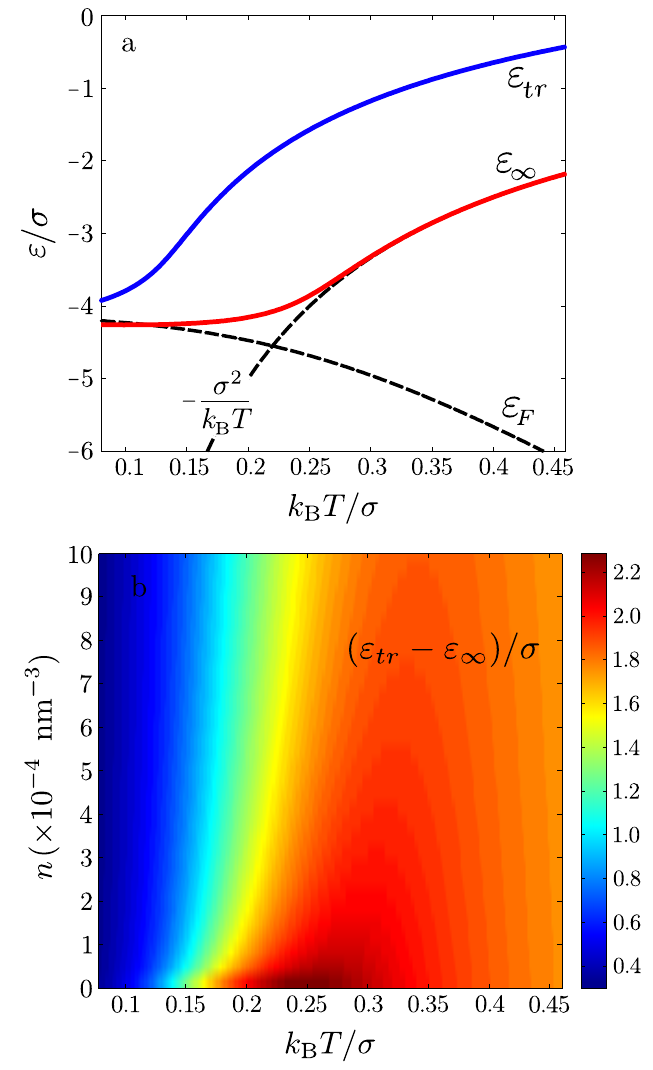}}
\caption{ \label{fig:2}\small \baselineskip=0.59cm
(a) Transport energy level $\varepsilon_{tr}$, as a function of disorder normalized temperature. Data are obtained using Eq.(\ref{eq:TE}) with $\sigma=0.065\text{ eV}$, $N=1\text{ nm}^{-3}$, and $n/N=1.6\times10^{-5}$. $\varepsilon_F$ is the Fermi level and $\varepsilon_{\infty}$ is the thermal equilibrium energy, approaching $-\sigma^2/k_{\mathrm{B}}T$ at high temperatures. (b) Heat map of $\varepsilon_{tr}-\varepsilon_{\infty}$ for a broad range of temperatures and exciton densities.
}
\end{figure}

We take into account a Gaussian distribution of energy states $g(\varepsilon)=N/\sqrt{2\pi\sigma^2} \allowbreak  \exp(-\varepsilon^2/2\sigma^2)$, with $N$ and $\sigma$ the total density of states (DOS) and the width of the distribution, respectively. If the relaxation process is completed during the lifetime $\tau$, \xcs will occupy states around the equilibrium energy $\varepsilon_\infty$ (see Fig.~\ref{fig:1}) at which the product $g(\varepsilon)f(\varepsilon, \varepsilon_F)$ maximizes~\cite{baranovskii2014theoretical}. Here, $f(\varepsilon, \varepsilon_F) = \{1+\exp[(\varepsilon-\varepsilon_F)/k_{\mathrm{B}}T]\}^{-1}$ is the Fermi distribution and $\varepsilon_F$ is the Fermi level, determined by the number density $n$ of the excitons as 
\begin{eqnarray}
n = \int_{-\infty}^{+\infty} g(\varepsilon)f(\varepsilon, \varepsilon_F) \mathrm{d}\varepsilon.
\label{eq:density}
\end{eqnarray}
Note that at low densities, the equilibrium energy $\varepsilon_\infty$ can be approximated by either $-\sigma^2/k_{\mathrm{B}}T$, at high temperatures~\cite{bassler1993charge}, or by $\varepsilon_F$, at low temperatures; see Fig.~\ref{fig:2}(a).

Now, we examine the possibility of the existence of a TE level $\varepsilon_{tr}$ in the energy distribution that can serve as the mobility edge in our exciton diffusion problem~\cite{BaranovskiiForster}. In the presence of such an energy level, \xcs with $\varepsilon>\varepsilon_{tr}$, will, on average, move downward in the distribution, towards the TE level. On the other hand, upward jumps of \xcs with $\varepsilon<\varepsilon_{tr}$ will be in the vicinity of $\varepsilon_{tr}$. 
If we express the mean jump distance as
\begin{eqnarray}
R_{\varepsilon_{tr}} = \left[
\frac{4\pi}{3}
\int_{-\infty}^{\varepsilon_{tr}} g(\varepsilon) f'(\varepsilon, \varepsilon_F) \mathrm{d}\varepsilon
\right]^{-{1}/{3}},
\label{eq:meanR}
\end{eqnarray}
we can obtain the following equation governing the position of the TE level for the \FO transport problem
\begin{eqnarray}
g(\varepsilon_{tr})f'(\varepsilon_{tr}, \varepsilon_F)  = \frac{1}{2k_{\mathrm{B}}T}
\int_{-\infty}^{\varepsilon_{tr}} g(\varepsilon)f'(\varepsilon, \varepsilon_F) \mathrm{d}\varepsilon
\label{eq:TE}
\end{eqnarray}
where $f'(\varepsilon, \varepsilon_F) = 1-f(\varepsilon, \varepsilon_F)$.  We have used the approach of Ref.\cite{baranovskii1995concept} to obtain the above equation, according to which one can find $\varepsilon_{tr}$ by maximizing the upward transfer rate; see Appendix \ref{sec:APP1} for more details. We emphasize that the  form of Eq.(\ref{eq:TE}) directly follows from the inverse \textit{sixth} power distance dependence of the dipole-dipole interaction. Eq.(\ref{eq:TE}) also shows that the position of $\varepsilon_{tr}$ is independent of the characteristic length $R_{\mathrm{F}}$ and the density $N$, in contrast to the charge transport problem in which $\varepsilon_{tr} = \varepsilon_{tr}(\alpha,N)$. Interestingly, Eq.(\ref{eq:TE}) does not acquire a solution for an exponential DOS. Again, this is in contrast to the charge transport problem, where for both Gaussian and exponential DOS one can find a TE level in the energy distribution. Charge transport in disordered semiconductors occurs via short-range transfer mechanism, with a rate similar to Eq.(\ref{eq:Forster}), but with $S(R) = \exp(- 2R/\alpha)$,  where $\alpha$ is the carrier localization length.

Fig.~\ref{fig:2}(a) illustrates $\varepsilon_{tr}$ as a function of  disorder normalized thermal energy. At high temperatures, $\varepsilon_{tr}$ lies near the center of the energy distribution. At lower temperatures, $\varepsilon_{tr}$ decreases to lower energies because by decreasing the temperature thermal activation to higher energies becomes less probable. We point out that a meaningful application of the TE level requires that the condition $\varepsilon_{tr}>\varepsilon_\infty$~\cite{baranovskii2014theoretical} is satisfied. To test this condition, we plot a heat map of $\varepsilon_{tr}-\varepsilon_\infty$ as a function of $k_{\mathrm{B}}T/\sigma$ and excitation density in Fig.~\ref{fig:2}(b), which shows that this condition is fulfilled over a broad range of temperatures and \xc densities. Thus the concept of the TE can be used for F\"{o}rster-type exciton transport. In what follows, we consider the weak excitation condition, with $n/N\ll 1$ and therefore $f'\approx1$. More precisely, we use $\sigma=0.065 \text{ eV}$, $N=1\text{ nm}^{-3}$ and $n/N=1.6\times10^{-5}$, corresponding to one \xc in a lattice of size $(40\text{ nm})^3$, as implemented in our kMC simulations. The same parameters were used in Fig.~\ref{fig:2}(a).

\section{\label{sec:Non}Non-equilibrium \xc dynamics} 
\subsection{\label{sec:dmrLevel}Demarcation energy level and energy relaxation} 
Having established the validity of the TE concept for exciton transport, let us now turn our attention to the main problem, that is, the description of the relaxation dynamics. Excitons, generated randomly in the DOS, progressively thermalize into deeper energies. Notwithstanding their way to the deep energy levels, \xcs need to be first activated to shallower energies, because the density of such levels is high in the energy distribution. Using the concept of the TE level we can say that these intermediate activations, necessary to approach thermal equilibrium, are most probable at the vicinity of the level $\varepsilon_{tr}$.  As first introduced by Tiedje and Rose~\cite{tiedje1981physical}, we can define a demarcation energy $\varepsilon_{m}(t)$ in the system, such that during time $t$ following the initial excitation, only the levels with $\varepsilon>\varepsilon_{m}(t)$ are likely to release their excitons to the TE level. Mathematically, this means that $t \nu(\varepsilon_m\rightarrow\varepsilon_{tr}) = \theta $, with $\theta$ being $\mathcal{O}(1)$. In a more explicit form, 
\begin{eqnarray}
t \frac{1}{\tau} \left(\frac{R_{\mathrm{F}}}{R_{\varepsilon_{tr}}} \right)^6
 \exp\left[- \frac{\varepsilon_{tr} - \varepsilon_m(t)}{k_{\mathrm{B}}T} \right] = \theta.
\label{eq:dm2}
\end{eqnarray}
From the above equation we find
\begin{eqnarray}
\varepsilon_m(t) = \varepsilon_{tr} - k_{\mathrm{B}}T\ln\left[ \frac{t}{\theta\tau}  
\left(\frac{R_{\mathrm{F}}}{R_{\varepsilon_{tr}}} \right)^6  \right]
\label{eq:dm3}
\end{eqnarray}
On the other hand, if we consider the low density condition, we can obtain the following equation for the mean jump distance from Eqs.(\ref{eq:meanR}) and (\ref{eq:TE})
\begin{eqnarray}
\frac{1}{R_{\varepsilon_{tr}}^3} = \frac{8\pi}{3}g(\varepsilon_{tr}) k_{\mathrm{B}}T
\label{eq:combine}
\end{eqnarray}
Inserting Eq.(\ref{eq:combine}) in Eq.(\ref{eq:dm3}), and using $g(\varepsilon)=N/\sqrt{2\pi\sigma^2}\exp(-\varepsilon^2/2\sigma^2)$, we get the following expression for the demarcation level
\begin{eqnarray}
\varepsilon_m(t) =
\varepsilon_{tr}\left(1+\frac{\varepsilon_{tr}}{\sigma^2/k_{\mathrm{B}}T}\right)
-k_{\mathrm{B}}T
\ln\left[ 
\left(N_{\mathrm{F}}\frac{k_{\mathrm{B}}T}{\sigma}\right)^2
\frac{2}{\theta\pi}\frac{t}{\tau}
\right]
\label{eq:dmFinal}
\end{eqnarray}
where $N_{\mathrm{F}}=(4\pi/3)R_{\mathrm{F}}^3N$.

According to Eq.(\ref{eq:dmFinal}), in the course of time, the demarcation level sinks to deeper energies. However, we note that this can continue only until time $t=\tau$, which is the intrinsic lifetime of the \x. If we interpret the demarcation energy as a quasi-Fermi level~\cite{bisquert2005interpretation}, at time $\tau$ most \xcs are accumulated around an energy level at which the product $g(\varepsilon)f[\varepsilon,\varepsilon_{m}(\tau)]$ maximizes. This energy is in fact the same energy $\varepsilon_{\tau}$ shown in Fig.~\ref{fig:1}. $\varepsilon_{\tau}$ is in general different from $\varepsilon_\infty$, but if the thermalization is completed during the \xc lifetime, we obtain $\varepsilon_{\tau}=\varepsilon_\infty$. The energy, $\varepsilon_{\tau}$ is experimentally available through fluorescence spectroscopy. We stress that our model can also be applied for exciton transport in the presence of quenching centers~\cite{MikhnenkoTrap,JPCCtrap}. In such situation one has to consider the demarcation energy at time $t<\tau$.

\begin{figure}[!t]
\scalebox{1}{\includegraphics{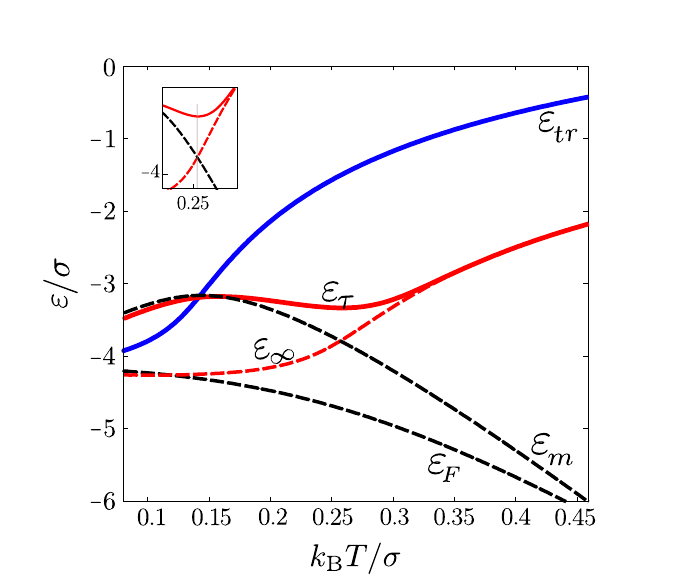}}
\caption{ \label{fig:3}\small \baselineskip=0.59cm
$\varepsilon_{tr}$ (TE level), $\varepsilon_{\tau}$ (energy relaxation during the \xc lifetime), $\varepsilon_{m}(\tau)$ (quasi-Fermi level), $\varepsilon_\infty$ (thermal equilibrium energy), and $\varepsilon_F$ (equilibrium Fermi level), as a function of disorder normalized temperature. Data are calculated using Eq.(\ref{eq:TE}) and (\ref{eq:dmFinal}) for $R_{\mathrm{F}}=5\text{ nm}$.
}
\end{figure}

The five energy levels discussed here, $\varepsilon_{tr}$ (transport energy level), $\varepsilon_{\tau}$ (energy relaxation during \xc lifetime), $\varepsilon_{m}(\tau)$ (demarcation or quasi-Fermi level at time $t=\tau$), $\varepsilon_\infty$ (thermal equilibrium energy), and $\varepsilon_F$ (equilibrium Fermi level), are plotted in Fig.~\ref{fig:3} for $R_{\mathrm{F}}=5\text{ nm}$. We have used $\theta\approx0.2$ since it gives excellent agreement with kinetic Monte Carlo (kMC) simulations, see below. As expected, at high disorder normalized temperatures the thermalization is nearly complete, and therefore $\varepsilon_{\tau}$ coincides with $\varepsilon_\infty$. However, by decreasing $k_{\mathrm{B}}T/\sigma$, $\varepsilon_{\tau}$ deviates from $\varepsilon_{\infty}$, owing to the incomplete thermalization during the \xc lifetime.  

Two temperature regions in Fig.~\ref{fig:3} need to be discussed in detail: 
 i) Region with $\varepsilon_{m}(\tau)>\varepsilon_{\infty}$. The relaxation energy $\varepsilon_{\tau}$ in this region reaches a minimum at a critical temperature where $\varepsilon_{m}(\tau)\approx\varepsilon_{\infty}$, and then \textit{increases} by decreasing the temperature, see inset of Fig.~\ref{fig:3}. This behavior, usually assigned to frustrated relaxation, has been observed experimentally for triplet excitons~\cite{hoffmann2010spectral}, and has been predicted through kMC simulations to occur also for \FO energy transfer~\cite{athanasopoulos2013Hoffmann}. Here, we see that our model can naturally produce the frustrated relaxation feature. 
ii) Region with $\varepsilon_m(\tau)>\varepsilon_{tr}$. In the temperature region given by the above condition, the multiple-trapping model is not applicable at all and introducing $\varepsilon_m(\tau)$ is physically meaningless. In this region, \xcs created in the system move, on average, downward towards the TE energy level, and therefore the picture of activation to a TE level is not correct. As we discuss below in the kMC section, an agreement between theory and simulation is not expected in this temperature region.

An important feature of the F\"{o}rster-type transport mechanism is that the \xc transfer rate is coupled to the spontaneous decay rate, $1/\tau$; see Eq.(\ref{eq:Forster}). Therefore, a longer \xc lifetime does not result in a higher degree of the thermalization, because the transfer rate, that determines the degree of thermalization, is also reduced. As a consequence, as predicted by Eq.(\ref{eq:dmFinal}) the demarcation energy at time $t=\tau$, and hence $\varepsilon_{\tau}$, are independent of the \xc lifetime.  On the other hand, $\varepsilon_{\tau}$ is a strong function of the \FO radius. We discuss this dependency below, when presenting the kMC simulation results.

\subsection{\label{sec:DL}Exciton diffusion length} 
An important physical quantity related to \xc transport is the diffusion length. In what follows we derive an expression for the \xc diffusion length using the TE level concept.  Since the diffusion length is given by\cite{ansari2012simulation}
\begin{eqnarray}
L_D=\sqrt{D\tau},
\label{eq:Length1}
\end{eqnarray}
we must first find the diffusion coefficient $D$. To obtain this, one can use \cite{baranovskii2000charge}
\begin{eqnarray}
D\approx R_{\varepsilon_{tr}}^2/\langle t \rangle
\label{eq:D}
\end{eqnarray}
where $\langle t \rangle$ is the mean time that \xcs spend in an energy state before activation to the TE level. $\langle t \rangle$ can be obtained by averaging the quantity $1/\nu(\varepsilon\rightarrow\varepsilon_{tr})$ for energies smaller than $\varepsilon_{tr}$~\cite{baranovskii2000charge,oelerich2012find}:
\begin{eqnarray}
\langle t \rangle = \tau \left( \frac{R_{\varepsilon_{tr}}}{R_{\mathrm{F}}} \right)^{\!\! 6}
\frac
{ \displaystyle \int _{-\infty}^{\varepsilon_{tr}}  
\exp\left(\frac{\varepsilon_{tr}-\varepsilon}{k_{\mathrm{B}}T} \right)  g(\varepsilon)f'[\varepsilon, \varepsilon_m(\tau)] \mathrm{d}\varepsilon}
{ \displaystyle \int _{-\infty}^{\varepsilon_{tr}}  
  g(\varepsilon)f'[\varepsilon, \varepsilon_m(\tau)] \mathrm{d}\varepsilon}
\label{eq:t average}
\end{eqnarray}
Combining Eqs.(\ref{eq:Length1}-\ref{eq:t average}), as shown in Appendix \ref{sec:APP2}, we get the following expression for the diffusion length
\begin{eqnarray}
L_D\approx
\left(
\frac{9\theta^3}{16\pi^2}
\,
\frac{N'-n'}{n'{^{3}}}
\right)^{1/6}
\label{eq:Length2}
\end{eqnarray}
where 
$n'=\int_{-\infty}^{\varepsilon_{tr}} g(\varepsilon)f[\varepsilon, \varepsilon_m(\tau)] \mathrm{d}\varepsilon$ and
$N'=\int_{-\infty}^{\varepsilon_{tr}} g(\varepsilon) \mathrm{d}\varepsilon$.
Note that, since according to Eq.(\ref{eq:dmFinal}) $\varepsilon_m(\tau)$ is a function of the \FO radius $R_{\mathrm{F}}$, the diffusion length is also $R_{\mathrm{F}}$ dependent. However, it is clear from Eq.(\ref{eq:Length2}) that the dependency of $L_D$ on $R_{\mathrm{F}}$ is more complex than that traditionally expected, that is, $L_{D}\propto \sqrt{D}\propto\sqrt{\nu}\sim R_{\mathrm{F}}^3$. This is because, for the problem of exciton transport in energetically disordered systems, $R_{\mathrm{F}}$ is not merely a  multiplicative factor, but according to Eq.(\ref{eq:dmFinal}), it also controls the thermalization process, which in turn, affects the dispersivity of the diffusion process. Another important result of our theory, as discussed in Appendix \ref{sec:APP3}, is that both the quantity $\varepsilon_{\tau}/\sigma$ and the diffusion length $L_D$ in Eq.(\ref{eq:Length2}) scale with the dimensionless disorder strength $\sigma/k_{\mathrm{B}}T$. Indeed, the scaling of both the exciton diffusion length and spectral relaxation has been predicted in the past by one of the authors using Monte Carlo simulations\cite{athanasopoulos2009PRB,SPIE,HowCharge}. In the following section, we test the validity of our approach to the problem of non-equilibrium \xc transport against Monte Carlo simulations.

\subsection{\label{sec:kMC}Kinetic Monte Carlo simulations} 
Monte Carlo simulations provide a insightful and predictive computational method for studying incoherent hopping transport phenomena in disordered semiconductors. In this manuscript we use a kinetic Monte Carlo method~\cite{athanasopoulos2009PRB} to simulate the time evolution of singlet exciton transport, confirm the validity of the developed theoretical model and test its applicability range. The computational protocol is as follows.

We consider a regular cubic cell of $\SI{40}{\nm} \times \SI{40}{\nm} \times \SI{40}{\nm}$ with a lattice constant $a=$\SI{1}{\nm}. Each lattice point corresponds to an exciton transport site, while periodic boundary conditions are implemented along all directions of the cell using the minimum image criterion. Individual Monte Carlo runs start by placing one exciton at a random site in the cell with each site having an energy drawn from a Gaussian distribution with a zero mean and variance $\sigma^2$. \FO transfer rates $\nu_{ij}$ from the exciton occupied site $i$ to each neighboring hopping site $j$, within a cut-off radius of $r_{\text{cut}}=\SI{5}{\nm}$, are calculated using Eq.\ref{eq:Forster}. At each Monte Carlo step, waiting times for each hopping event are calculated according to $\tau_{ij}=-\frac{1}{\nu_{ij}}\ln X$, with $X$ a random number from a box distribution from zero to unity, resulting to 514 events for the chosen cut-off radius. An additional waiting time for exciton recombination is computed as $\tau_{ir}=-\tau\ln X$. If the event with the shorter waiting time is a hopping event, then exciton transfers to the new site and simulation advances whereas if it is recombination, the exciton is removed from the system and the run is terminated. By averaging over $10^5$ individual exciton trajectories we obtain the quantities of interest, ie the relaxation energy $\varepsilon_{\tau}$ and the diffusion length $L_D$. The first is calculated from the final energy of each exciton before recombination, while the latter from the displacement between the initial, exciton generation, and the final, exciton recombination, position. We allow to vary independently the temperature T and \FO radius $R_F$ parameters, while disorder $\sigma$ and lifetime $\tau$ remain constant. In fact, due to the \FO rate inverse dependence on $\tau$, $\tau$ does not impact neither the $\varepsilon_{\tau}$ nor the $L_D$ values, while a scaling law exists for both of them with respect to the dimensionless disorder parameter $\sigma/k_{\mathrm{B}}T$~\cite{athanasopoulos2009PRB,athanasopoulos2013Hoffmann}.

\begin{figure}[!t]
\scalebox{1}{\includegraphics{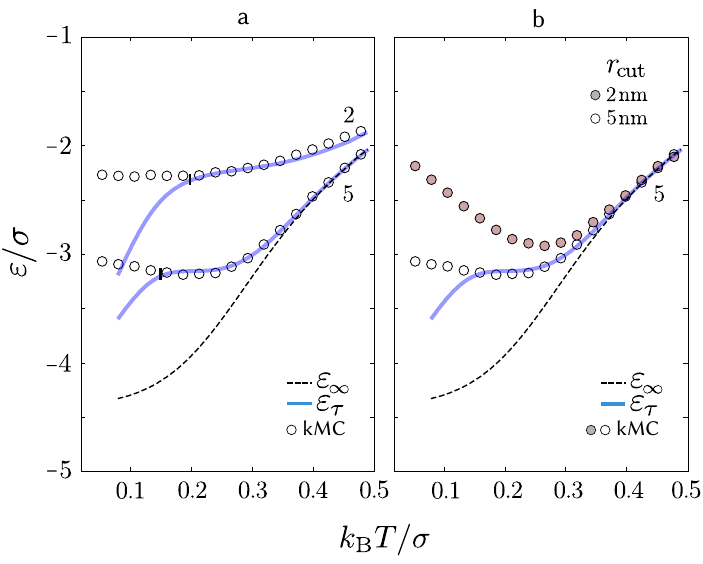}}
\caption{ \label{fig:4}\small \baselineskip=0.59cm
(a) Energy relaxation during the \xc lifetime, $\varepsilon_{\tau}$ as a function of disorder normalized temperature. kMC simulations (circles) and theory (solid lines), for two different \FO radii, $R_{\mathrm{F}}=2$ and $5 \text{ nm}$. The critical points at which $\varepsilon_m(\tau)=\varepsilon_{tr}$, are indicated as segments. Dashed line indicate the thermal equilibrium energy $\varepsilon_\infty$. 
(b) Same as (a) with $r_{\mathrm{cut}}=5 \text{nm}$ (empty circles) and 2 nm (filled circles).  
}
\end{figure} 

\begin{figure}[!t]
\scalebox{1}{\includegraphics{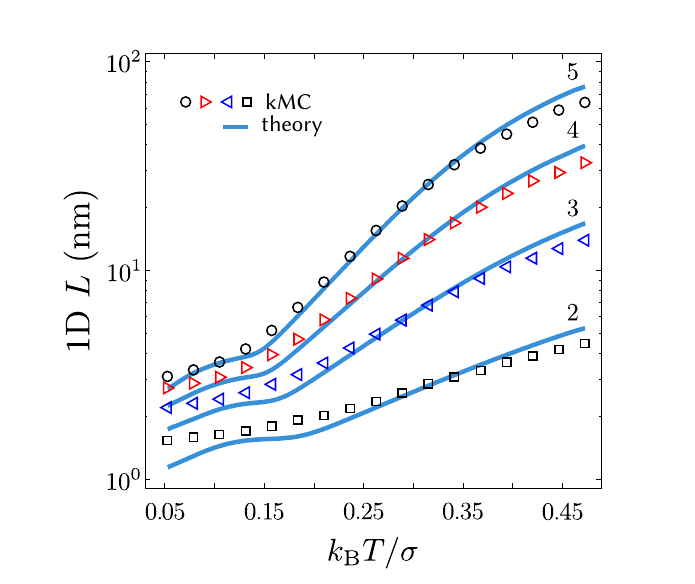}}
\caption{ \label{fig:5}\small \baselineskip=0.59cm
Diffusion length $L_D$ as a function of disorder normalized temperature. Data from kMC simulation (symbols) and theory (solid lines), for different \FO radii, $R_{\mathrm{F}}=2$, $3$, $4$ and $5\text{ nm}$. }
\end{figure}

The central results comparing theory with Monte Carlo simulations are presented in Fig.~\ref{fig:4} for the spectral relaxation and Fig.~\ref{fig:5} for the diffusion length. Figure~\ref{fig:4} (a) shows the Monte Carlo results for $\varepsilon_{\tau}$, for two \FO radii $R_{\mathrm{F}}=2$ and $5 \text{ nm}$. The theoretical predictions, calculated based on the TE concept and using the averaging method (see Appendix \ref{sec:APP4}), are also shown in the figure. As pointed out above in part A, the multiple-trapping picture is not valid when $\varepsilon_m(\tau)>\varepsilon_{tr}$. The exact points at which $\varepsilon_m(\tau)=\varepsilon_{tr}$ are calculated and marked in the figure. In the region where the TE concept is applicable, the theory is in very good agreement with the kMC results. Since the density of the energy levels is higher near the center of the Gaussian distribution, most \xcs generated in the system will have energies $\varepsilon\approx0$ and according to the TE concept, those \xcs initially move, \textit{on average}, downwards to the TE level $\varepsilon_{tr}$. However, en-route to the TE level some upward in energy jumps are also necessary to avoid the blockade of excitons due to disorder. Therefore, for larger \FO radii, the TE concept is valid over a broader range of temperatures, because a larger $R_{\mathrm{F}}$ results to a higher probability to overcome local energy barriers.
A recent combined experimental and computational study highlighted the dominant contribution of long-distance jumps to singlet \xc migration in metal-organic frameworks~\cite{JACSBeyond}. To illustrate the importance of long-distance hopping, we have also performed simulations with $r_{\mathrm{cut}}=2 \text{ nm}$ (i.e. restricting exciton hopping only to the first 32 nearest neighbors). Fig.~\ref{fig:4}(b) shows that in comparison to $r_{\mathrm{cut}}=5 \text{ nm}$ (514 nearest neighbors), the energy relaxation shows a pronounced frustrated dynamics, inconsistent with the theory prediction. This clearly demonstrates that especially at low temperatures, long-range jumps contribute significantly to the relaxation process.  In other words, due to the long-range nature of the \FO mechanism, modeling the singlet \xc transport as a simple nearest neighbor random walk process may result in an incorrect description of the energy transfer dynamics. We can also conclude that for inherently short-range transport mechanism, like charge or triplet \xc transport problem, a strong frustration is expected, as indeed reported in earlier simulations~\cite{athanasopoulos2013Hoffmann,HowCharge}.

\begin{figure}[!t]
\scalebox{1}{\includegraphics{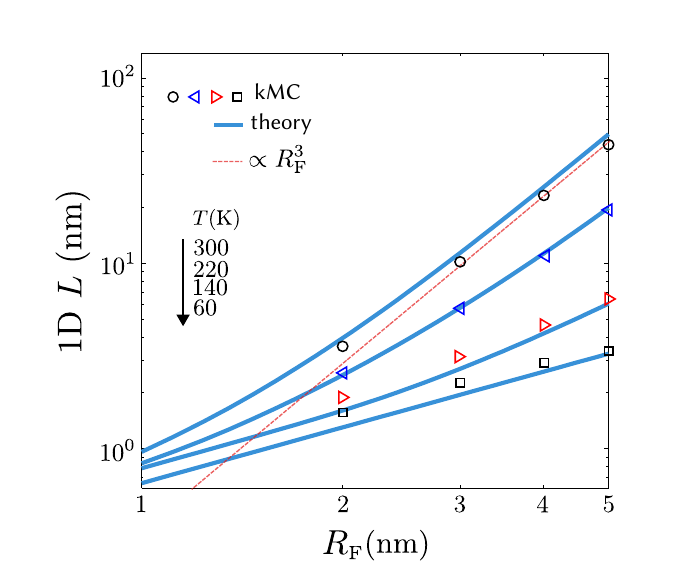}}
\caption{ \label{fig:6}\small \baselineskip=0.59cm
Diffusion length $L_D$ as a function of the \FO radius for different temperatures on a log-log scale. Data from kMC simulations (symbols) and theory (solid lines). Dashed line indicates the slope expected from $L_{D}\propto R_{\text{F}}^3$.} 
\end{figure}

Fig.~\ref{fig:5} compares $L_D$ obtained from the kMC simulations with those calculated using Eq.(\ref{eq:Length2}). Apart from an additional constant factor ($\approx1.5$) needed to fit the theory to the simulation (see Section \ref{sec:Subdiff}), the theoretical results are in good agreement with the kMC simulations showing a steep increase of the diffusion length with disorder normalized thermal energy. We point out that in contrast to spectral relaxation, reliable estimates for $L_D$ from the theoretical model can be obtained even in the regime where $\varepsilon_m(\tau)<\varepsilon_{tr}$ as $L_D$ is less sensitive to $\varepsilon_m(\tau)$ in that region. It must be noted that our results are in agreement with experimental reports on the temperature dependence of the exciton diffusion length~\cite{mikhnenko2008temperature,lin2015temperature}. Finally, Fig.~\ref{fig:6} shows that the traditional picture of $L_{D}\propto R_{\text{F}}^3$ does not hold true at the intermediate and low temperature region, as predicted and discussed in the theory section above.

\begin{figure}[!t]
\scalebox{1}{\includegraphics{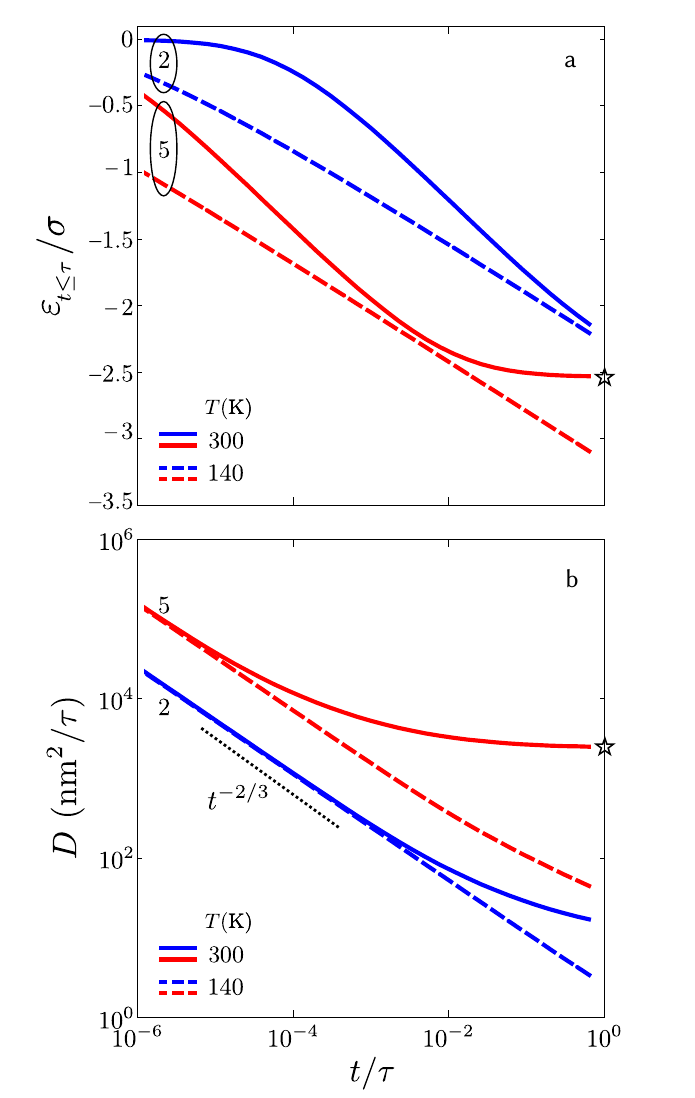}}
\caption{ \label{fig:7}\small \baselineskip=0.59cm
(a) Energy relaxation (shift) as a function of time. Data for $T=140$ and $300 \text{ }\mathrm{K}$ and $R_F=2$ and $5 \text{ nm}$. 
(b) Diffusion coefficient $D(t)$ as a function of time calculated using Eq.(\ref{eq:diffCoeff2}). Data for $T=140$ and $300 \text{ }\mathrm{K}$ and $R_F=2$ and $5 \text{ nm}$. The dotted line shows the scaling of the diffusion coefficient with time in the non-equilibrium regime. Stars indicate the equilibrium energies. 
}
\end{figure}
\subsection{\label{sec:Subdiff}Subdiffusive Transport}
Having established the effectiveness of the analytical model to describe spectral relaxation, we now turn our attention to obtaining the time dynamics of exciton diffusion. Very recently, it has been reported experimentally that exciton diffusion in a system of disordered colloidal quantum dots is dispersive and can be described as a subdiffusive transport~\cite{akselrod2014subdiffusive}, in which $D(t)\propto t^{\beta}$ with $\beta<0$. Similar results have been obtained from Monte Carlo simulations for triplet exciton and charge transport in a Gaussian DOS~\cite{HowCharge, schnonherr1980dispersive}. To investigate whether our model can explain these observations, we expand the TE concept to take into account the time dependence of the dynamics for $t<\tau$. This can be achieved by considering the demarcation energy $\varepsilon_m(t)$, instead of $\varepsilon_m(\tau)$ used in the previous calculations. Below, we present results for the energy relaxation shift $\varepsilon_{t\leq \tau}$ and the diffusion coefficient $D(t)$, while we derive the time-dependent expressions in Appendix \ref{sec:APP2}. 

Fig.~\ref{fig:7} shows the temporal evolution of $\varepsilon_{t\leq \tau}$ and $D(t)$ for two different temperatures and two \FO radii, $R_{\mathrm{F}}=2$ and $ 5 \text{ nm}$. As seen in Fig.~\ref{fig:7}(a), in the course of time, \xcs relax to lower energy levels. For a \FO radius of $R_{\mathrm{F}}= 5 \text{ nm} $ and at high temperatures, \xcs reach the equilibrium energy during their lifetime (this is also apparent in Fig.~\ref{fig:4}(a)) and a stationary state is indeed established at $t<\tau$. In contrast, at low temperatures and/or small \FO radius, the relaxation process is incomplete and the stationary state can not obtained. Interestingly, our theoretical results for low temperatures show a linear dependence with time in the logarithmic scale $\varepsilon_{t\leq\tau}\sim - \ln(t/\tau) $, with the same slope for both $R_{\mathrm{F}}=2$ and $5\text{ nm}$, and are in agreement with results from Movaghar \textit{et. al.}~\cite{movaghar1986diffusion}.
The time evolution of the diffusivity is shown in Fig.~\ref{fig:7}(b). As a result of \xc relaxation to lower energy levels with time, based on the multiple-trapping picture the waiting time needed to jump to the TE level increases with time. Therefore, the diffusion coefficient becomes time-dependent, ie dispersive transport, and decreases with time. As derived in Appendix \ref{sec:APP2}, in this non-equilibrium regime we obtain
\begin{eqnarray}
D(t) \propto \left(\frac{t}{\tau}\right) ^{-2/3}
\label{eq:Dvst}
\end{eqnarray}
which clearly demonstrates the dispersive nature of singlet exciton diffusion. Nevertheless, at high temperatures and large \FO radius, since equilibrium can be established during the exciton lifetime, the diffusion coefficient approaches its equilibrium, time-independent, value. 

In obtaining Eq.(\ref{eq:Length2}) for the diffusion length, the diffusion coefficient at time $t=\tau$ has been used in the calculations. However, since the \xc transport occurs  almost entirely in the non-equilibrium regime and the diffusion coefficient is time-dependent, using $D(t=\tau)$ may result in an underestimation of the diffusion length. This argument shows why an additional factor was required to fit the theory with the kMC results in Fig.~\ref{fig:5}. One can estimate this factor by using the following relation for the diffusion length of  \xcs 
\begin{eqnarray}
L_D = \sqrt{ \int _{0}^{\tau} D(t) \mathrm{d}t }
\label{eq:LwithDvst}
\end{eqnarray}
Using Eq.(\ref{eq:Dvst}) we have
\begin{eqnarray}
L_D\approx\sqrt{D(\tau)\tau \int _{0}^{\tau} \left({t/\tau}\right) ^{-2/3} \mathrm{d}\left({t/\tau}\right) }=
\sqrt{3}\times\sqrt{D(\tau)\tau}
\label{eq:LwithDvst2}
\end{eqnarray}
The factor $\sqrt{3}$ justifies the additional factor used in Fig.~\ref{fig:5} to match the theory with the kMC results.

\section{\label{CONCL} Conclusion}
A theory for singlet exciton hopping transport has been developed and tested. It describes diffusive transport via long-range \FO transfer in a Gaussian distribution of localized states through a multiple-trapping mechanism, with the TE playing the role of the mobility edge. The theory provided in this paper fully describes the transition from equilibrium to non-equilibrium transport. The global validity range of the theory is illustrated by comparison to Monte Carlo simulations. We find that for \FO radius values smaller than 5 nm, typical in organic semiconductors, exciton transport occurs mainly in the non-equilibrium regime and the diffusion length deviates from the cubic dependence upon the \FO radius. An important feature of the theory is that it takes into account explicitly the temporal evolution of the spectral relaxation energy and diffusivity and can be used to understand time-gated spectroscopic experiments in a wide-range of disordered semiconducting materials. Understanding the exciton dynamics is also important for exploiting novel device applications. In the current work we take a step towards this goal and anticipate that it will motivate further studies. In future work we hope to tackle the excitation density dependence of the relaxation dynamics and transport in spatially correlated disordered systems.  

\acknowledgments This project has received funding from the Universidad Carlos III de Madrid, the European Union's Seventh Framework Programme for research, technological development and demonstration under grant agreement n$^{\text{o}}$ 600371, el Ministerio de Econom{\'i}a, Industria y Competitividad (COFUND2014-51509), el Ministerio de Educaci{\'o}n, cultura y Deporte (CEI-15-17) and Banco Santander.

\appendix
\section{\label{sec:APP1} Position of the Transport Energy}
According to Eq.(\ref{eq:Forster}), the upward \xc jump rate is given by 
\begin{eqnarray}
\nu(\varepsilon_d\rightarrow \varepsilon_a) = \frac{1}{\tau} 
\left( \frac{R_{\mathrm{F}}}{R} \right)^6  \exp\left(- \frac{\varepsilon_a - \varepsilon_d }{k_{\mathrm{B}}T} \right)
\label{eq:ForsterAPP}
\end{eqnarray}
where $\varepsilon_a - \varepsilon_d > 0$ is the difference between the acceptor and donor energy. Let us denote this rate by $\nu_{\uparrow}(\varepsilon_d,\varepsilon_a, R)$. For steep energy distributions, the typical upward jump distance is given by [Eq.(\ref{eq:meanR}) in the main text]
\begin{eqnarray}
R_{\varepsilon_{a}} = \left[
\frac{4\pi}{3}
\int_{-\infty}^{\varepsilon_a} g(\varepsilon) f'(\varepsilon, \varepsilon_F) \mathrm{d}\varepsilon
\right]^{-{1}/{3}}
\label{eq:meanRAPP}
\end{eqnarray}
Now, according to the standard approach of calculating the transport energy level, we seek to find if such an acceptor energy level exists that it maximizes all typical upward jumps, independent of the donor energy. In other words, we look for a unique acceptor energy, $\varepsilon_{tr}$, that meets the condition
\begin{eqnarray}
\left.\frac{\partial \nu_{\uparrow}(\varepsilon_d,\varepsilon_a,R_{\varepsilon_a})}
{\partial \varepsilon_a}\right\vert_{\varepsilon_a=\varepsilon_{tr}} = 0.
\label{eq:conditionApp}
\end{eqnarray}
By algebraic manipulation of the above equation we obtain Eq.(\ref{eq:TE}).

\section{\label{sec:APP2} Diffusion coefficient}
In this appendix a general expression for the diffusion coefficient is obtained, from which the time-dependency of the diffusion coefficient and the singlet diffusion length can be extracted. First, we note that the integral in the numerator of Eq.(\ref{eq:t average}) can be rewritten as 
\begin{eqnarray}
\exp\left(\frac{\varepsilon_{tr}-\varepsilon_m}{k_{\mathrm{B}}T} \right) 
 \int _{-\infty}^{\varepsilon_{tr}}  
\exp\left(\frac{\varepsilon_{m}-\varepsilon}{k_{\mathrm{B}}T} \right)  g(\varepsilon)f'(\varepsilon, \varepsilon_m) \mathrm{d}\varepsilon,
\label{eq:numerator1}
\end{eqnarray}
where for brevity we have used $\varepsilon_m$ for $\varepsilon_m(t)$. This, bearing in mind that $f'=1-f$, can be simplified as
\begin{eqnarray}
\exp\left(\frac{\varepsilon_{tr}-\varepsilon_m}{k_{\mathrm{B}}T} \right) 
\int _{-\infty}^{\varepsilon_{tr}}  
g(\varepsilon)f(\varepsilon, \varepsilon_m) \mathrm{d}\varepsilon.
\label{eq:numerator2}
\end{eqnarray}
On the other hand, using Eq.(\ref{eq:dm2}), for the exponential term in the above equation we have
\begin{eqnarray}
\exp\left(\frac{\varepsilon_{tr}-\varepsilon_m}{k_{\mathrm{B}}T} \right) =
\frac{t}{\tau\theta}\left(\frac{R_{\mathrm{F}}}{R_{\varepsilon_{tr}}} \right)^6
\label{eq:expNumerqtor}
\end{eqnarray}
Using these simplifications, and if we define $n'=\int_{-\infty}^{\varepsilon_{tr}} g(\varepsilon)f(\varepsilon, \varepsilon_m) \mathrm{d}\varepsilon$ and
$N'=\int_{-\infty}^{\varepsilon_{tr}} g(\varepsilon) \mathrm{d}\varepsilon$, we obtain
\begin{eqnarray}
D(t)=\frac{\theta}{t}R_{\varepsilon_{tr}}^2
\frac{N'-n'}{n'}
\label{eq:diffCoeff}
\end{eqnarray}
that, using Eq.(\ref{eq:meanR}),  can be rewritten as
\begin{eqnarray}
D(t)=\frac{\theta}{t} \left( \frac{4\pi}{3} \right)^{-2/3} 
\frac{\left(N'-n'\right)^{1/3}}{n'}
\label{eq:diffCoeff2}
\end{eqnarray}
From this general result, one can obtain Eq.(\ref{eq:Length2}) for the diffusion length $L_{D}=\sqrt{D(\tau)\tau}$. 

To obtain the time-evolution of the diffusion coefficient in non-equilibrium regime, we use the fact that the demarcation energy is high at short and intermediate times such that we can write $f\approx 1$  and $1-f\approx \exp\{[\varepsilon - \varepsilon_m(t)]/{k_{\mathrm{B}}T}\}$. Therefore, since  $\varepsilon_m(t) = \varepsilon_m(\tau) - {k_{\mathrm{B}}T} \ln(t/\tau)$, we can obtain the following time-dependent behavior for the diffusion coefficient (valid only for the non-equilibrium regime)
\begin{eqnarray}
D(t) \propto (t/\tau)^{-2/3}
\label{eq:diffCoeffTime}
\end{eqnarray}
On the other hand, at the equilibrium regime where the demarcation energy lies deep in the energy distribution, we can use the approximation $f\approx \exp\{-[\varepsilon - \varepsilon_m(t)]/{k_{\mathrm{B}}T}\}$ and $N'-n'\approx N'$. These approximations result in a stationary diffusion coefficient as
\begin{eqnarray}
D_{\mathrm{st}} \propto (t/\tau)^0
\label{eq:diffCoeffTimeInft}
\end{eqnarray}

\section{\label{sec:APP3} Scaling behavior of the diffusion length}
Eq.(\ref{eq:density}) shows that at a given density $n$, the Fermi level $\varepsilon_F$ is determined by the temperature and the width of the energy distribution. By expressing this integral in terms of a new variable $x=\varepsilon/\sigma$, we find that the temperature-normalized Fermi level, that is $\varepsilon_F/k_{\mathrm{B}}T$, is a function of the dimensionless disorder parameter $\sigma/k_{\mathrm{B}}T$. Using this result, and the same change-of-variable for the integral of Eq.(\ref{eq:TE}), we find that $\varepsilon_{tr}/\sigma$ is a function of $\sigma/k_{\mathrm{B}}T$. Inspection of Eq.(\ref{eq:dmFinal}) for the demarcation level shows that the same scaling behavior holds for $\varepsilon_m(\tau)/\sigma$, and since $\varepsilon_{\tau}$ is the energy at which the product $g(\varepsilon)f[\varepsilon,\varepsilon_{m}(\tau)]$ maximizes, we find that $\varepsilon_{\tau}/\sigma$ also scales with $\sigma/k_{\mathrm{B}}T$. Using the above scaling features and Eq.(\ref{eq:Length2}) we obtain that $L_{D}=L_{D}(\sigma/k_{\mathrm{B}}T)$.

\section{\label{sec:APP4} Averaging method for the calculation of the relaxation energy}
The equilibrium energy $\varepsilon_{\infty}$ can be calculated in two different ways. As pointed out in the main text, we have introduced $\varepsilon_{\infty}$ as the energy that maximizes the product $g(\varepsilon)f(\varepsilon,\varepsilon_{F})$. Accordingly, the relaxation energy $\varepsilon_{\tau}$ can be found by maximizing the product $g(\varepsilon)f[\varepsilon,\varepsilon_{m}(\tau)]$. On the other hand, one can define the equilibrium or relaxation energy as the average energy of the carriers. In this definition, the equilibrium energy is calculated as 
\begin{eqnarray}
\langle \varepsilon \rangle = 
\frac
{ \displaystyle \int \varepsilon g(\varepsilon)f(\varepsilon, \varepsilon_{F}) \mathrm{d}\varepsilon }
{ \displaystyle \int                  g(\varepsilon)f(\varepsilon, \varepsilon_{F}) \mathrm{d}\varepsilon }
\label{eq:AvgEq}
\end{eqnarray} 
To obtain $\varepsilon_{\tau}$, one needs to replace $\varepsilon_F$ with $\varepsilon_m(\tau)$ in the above equation. We find that the averaging method gives excellent agreement with Monte Carlo simulations. In comparison, the method of maximizing the product $f\times g$ results in slightly lower values for the equilibrium energy at intermediate and higher temperatures and a more pronounced minimum (Fig.~\ref{fig:3}). However, since the product $f\times g$ is approximately a symmetric function of energy, the two definitions result in the same overall trend and similar values for the relaxation energy. From a practical point of view, while the first method is numerically more tractable, the second definition is most suitable for comparing with kMC simulation results, where the relaxation energy is obtained by averaging over different exciton trajectories. Throughout this work we adopted the first method, except in Fig.~\ref{fig:4} where we compare $\varepsilon_{\tau}$ with kMC calculations.

\bibliography{excitRef}

\end{document}